
\input phyzzx
\hoffset=0.2truein
\voffset=0.1truein
\hsize=6truein
\def\TITLEPAGE{\frontpagetrue}
\def\IASSNS#1{\hbox to\hsize{\tenpoint \baselineskip=12pt
        \hfil\vtop{
        \hbox{\strut iassns-hep-93-#1}
}}}
\def\SUNY#1{\hbox to\hsize{\tenpoint \baselineskip=12pt
        \hfil\vtop{
        \hbox{\strut itp-sb-93-#1}
}}}

\def\IAS{
        \centerline{School of Natural Science, Institute for Advanced Study}
\centerline{Olden Lane, Princeton, New Jersey 08540}}
\def\SUNYSB{
        \centerline{Institute for Theoretical Physics, State %
University of New York}
\centerline{Stony Brook, New York 11794-3840}
}
\def\TITLE#1{\vskip .5in \centerline{\fourteenpoint #1}}
\def\AUTHOR#1{\vskip .2in \centerline{#1}}

\def\ABSTRACT#1{\vskip .2in \vfil \centerline{\twelvepoint
\bf Abstract}
   #1 \vfil}
\def\ENDTITLEPAGE{\vfil\eject\pageno=1}
\hfuzz=5pt
\tolerance=10000
\TITLEPAGE
\IASSNS{6}
\SUNY{60}
\TITLE{$SO(10)$ Cosmic Strings and $SU(3)_{color}$ Cheshire Charge}

\AUTHOR{Martin Bucher}

\IAS
\centerline{and}
\SUNYSB

\AUTHOR{Alfred S. Goldhaber}

\SUNYSB

{\baselineskip=14pt\ABSTRACT{
Certain cosmic strings that occur in GUT models such as $SO(10)$ can carry
a magnetic flux which acts nontrivially on objects carrying $SU(3)_{color}$
quantum numbers. We show that such strings are non-Abelian Alice strings
carrying nonlocalizable colored ``Cheshire" charge. We
examine claims made in the literature that $SO(10)$ strings
can have a long-range, topological Aharonov-Bohm interaction that
turns quarks into leptons, and observe that such a process is
impossible. We also discuss flux-flux scattering using a multi-sheeted
formalism.}}
\bigskip
\centerline{[PACS 11.15, 98.90.C, 12.10.D, 5.30.C]}
\bigskip\rightline{April 1993, Revised 8 October}
\ENDTITLEPAGE
\eject

\REF\kib{T. Kibble, G. Lazarides and Q. Shafi, ``Strings in $SO(10),$
Phys. Lett. {\bf B113,} 237 (1982)}
\REF\aryal{M. Aryal and A. Everett, ``Properties of $Z_2$ Strings,"
Phys. Rev. {\bf D35,} 3105 (1987)}
\REF\everetta{A. Everett, ``Cosmic Strings in Unified Gauge
Theories," Phys. Rev. {\bf D24,} 858 (1981).}
\REF\olive{D. Olive and N.Turok, ``$Z_2$ Vortex Strings in
Grand Unified Theories," Phys. Lett. {\bf B117,} 193 (1982)}
\REF\guta{E. Witten, ``Superconducting Strings,"
Nucl. Phys. {\bf B249,} 557 (1985)}
\REF\cosmic{See, for example, A. Vilenkin,
``Cosmic Strings and Domain Walls,"
Phys. Rep. {\bf 121,} 263 (1984); J. Preskill, ``Vortices and
Monopoles,'' in {\it Architecture of the Fundamental Interactions
at Short
Distances}, ed. P. Ramond and R. Stora (North-Holland, Amsterdam,
1987).}
\REF\aharonov{Y. Aharonov and D. Bohm, Phys. Rev. {\bf 115,} 485 (1959).}
\REF\yang{T. Wu and C.N. Yang, Phys. Rev. {\bf D12,} 3845 (1975).}
\REF\horvathy{P. Horvathy, ``The Nonabelian Aharonov-Bohm Effect,"
Phys. Rev. {\bf D33,} 407 (1986).
R. Sundrum and L. Tassie, ``Nonabelian Aharonov-Bohm Effects,
Feynman Paths, and Topology," J. Math. Phys. {\bf 27,} 1566 (1986).}
\REF\wal{M. Alford and F. Wilczek,
``Aharonov-Bohm Interaction of Cosmic Strings with Matter,"
Phys. Rev. Lett. {\bf 62,} 1071 (1989).}
\REF\preskill{J. Preskill and L. Krauss, Nucl. Phys. {\bf B341,}
50 (1990).}
\REF\alford{M. Alford, K. Benson, S. Coleman, J. March-Russell,
and F. Wilczek, ``Interactions and Excitations of Non-Abelian
Vortices," Phys. Rev. Lett. {\bf 64,} 1632 (1990) [Erratum: %
{\bf 65,} 668 (1990).];
M. Alford, K. Benson, S. Coleman, J. March-Russell,
and F. Wilczek, ``Zero Modes of  Non-Abelian
Vortices, Nucl. Phys. {\bf B349,} 414 (1991).}
\REF\ginsparg{S. Coleman and P. Ginsparg, unpublished (1982).}
\REF\bucher{M. Bucher, H.K. Lo, J. Preskill, ``Topological
Approach to Alice Electrodynamics," Nucl. Phys. {\bf B386,}
3 (1992).}
\REF\mbb{M. Bucher, K.M. Lee and J. Preskill, ``Detecting Discrete
Cheshire Charge," Nucl. Phys. {\bf B386,} 27 (1992).}
\REF\brand{L. Perivolaropoulos, A. Matheson, A. Davis,
and R. Brandenberger, ``Nonabelian Aharonov-Bohm Baryon
Decay Catalysis," Phys. Lett. {\bf B245,} 556 (1990).}
\REF\ma{C.P. Ma, ``$SO(10)$ Cosmic Strings and Baryon Number
Violation," MIT Preprint (1991).}
\REF\julia{B. Julia and A. Zee, ``Poles with Both Magnetic and Electric
Charge in Nonabelian Gauge Theory," Phys Rev. {\bf D11,} 2227 (1975).}
\REF\sky{T. Skyrme, Proc. Royal Soc. {\bf A260,} 127 (1961).}
\REF\schwarz{A.S. Schwarz, Nucl. Phys. {\bf B209,} 141 (1982).}
\REF\barut{A. Barut and R. Raczka, {\it Theory of Group Representations
and Applications,} (Singapore: World Scientific, 1986).}
\REF\everett{A. Everett, ``New Mechanism for String Superconductivity,"
Phys. Rev. Lett {\bf 61,} 1807 (1988).}
\REF\bala{A. Balachandran, F. Lizzi and V. Rodgers, ``Topological
Symmetry Breakdown in Nematics and He-3," Phys. Rev. Lett. {\bf 52,}
1818 (1984).}
\REF\wzb{M. Alford, J. March-Russell and F. Wilczek,
``Enhanced Baryon Number Violation
Due To Cosmic Strings," Nucl. Phys. {\bf B328,} 140 (1989).}
\REF\wilczekwu{F. Wilczek and Y. Wu, ``Spacetime Approach
to Holonomy Scattering," Phys. Rev. Lett. {\bf 65,} 13 (1990.)}
\REF\mba{M. Bucher, ``The Non-Abelian Aharonov-Bohm Effect and
Exotic Statistics for Nonabelian Vortices and Cosmic Strings, "
Nucl. Phys. B350, 193 (1991).}
\REF\peonaru{V. Poenaru and G. Toulouse, ``The Crossing of Defects in
Ordered Media and the Topology of 3-Manifolds," J. of Phys.
{\bf 8,} 887 (1977).}
\REF\bais{F. Bais, ``Flux Metamorphosis,"
Nucl. Phys. {\bf B170,} 32 (1980)}
\REF\baistwo{F. Bais, P. van Driel and M. de Wild Propitius,
``Quantum Symmetries in Discrete Gauge Theories,"
Phys. Lett. {\bf B280,} 63 (1992);
``Anyons in Discrete Gauge Theories with Chern-Simons Terms,"
Nucl. Phys. {\bf B393,} 547 (1993).}
\REF\coleman{M. Alford, S. Coleman, and J. March-Russell,
``Disentangling Nonabelian Discrete Quantum Hair," Nucl. Phys.
{\bf B351,}735 (1991)}
\REF\loandpreskill{H.K. Lo and J. Preskill, `` Non-Abelian Vortices and
Non-Abelian Statistics," Caltech Preprint CALT-68-1867 (1993).}
\REF\preskilltwo{M. Alford, K.M. Lee, J. March-Russell and J. Preskill,
``Quantum Field Theory of Nonabelian Strings and Vortices," Nicl. Phys.
{B384,} 251 (1992).}
\REF\peak{D. Peak and A. Inomata, J. Math Phys. {\bf 10,}
1422 (1969). See also D. Arovas, in A. Shapere and F. Wilczek, Eds.
{\it Geometric Phases in Physics,} (Singapore: World Scientific, 1989).}
\REF\sommerfeld{A. Sommerfeld,
Proc. London Math. Soc. {\bf 28,} 417 (1897).}
\REF\carslaw{H. Carslaw,``Diffraction of Waves by a Wedge of Any Angle,"
Proc. London Math. Soc. {\bf 8,} 365 (1910);
Proc. London Math. Soc. {\bf 18,} 291 (1919).  }
\REF\bornandwolf{M. Born and E. Wolf, ``Principles of Optics",
Sixth Edition, (Oxford: Pergamon Press, 1989)}
\REF\anyons{See for example, F. Wilczek, ``Fractional Statistics and Anyon
Superconductivity," (Singapore: World Scientific, 1990).}

\def\gtorder{\mathrel{\raise.3ex\hbox{$>$}\mkern-14mu
             \lower0.6ex\hbox{$\sim$}}}
\def\ltorder{\mathrel{\raise.3ex\hbox{$<$}\mkern-14mu
             \lower0.6ex\hbox{$\sim$}}}

\chapter{Introduction}

Cosmic strings are vortex lines that occur as the result of
spontaneous symmetry breaking in certain quantum field theories.
Although there are none in the minimal $SU(5)$ GUT,
such strings occur generically in many GUT models based
on larger groups, such as $SO(10)$ \refmark{\kib -\guta }. Cosmic strings
have been
proposed as seeds for structure formation in the universe,
and the approximate string tension required for the correct level of
density perturbations nicely coincides with the GUT scale suggested
by the unification of the coupling constants \refmark{\cosmic }.

This paper is about the Aharonov-Bohm interactions of GUT cosmic
strings. It has been noted that although the region surrounding
the cosmic string is pure vacuum (as long as the region
is simply-connected
and sufficiently distant from the string core), the magnetic flux
confined in the core of the string can give rise to an Aharonov-Bohm
interaction \refmark{\aharonov ,\yang ,\horvathy ,\wal }.
This long-range interaction is
of a topological nature, and completely determined by
the magnetic flux carried by the string, which we define in
terms of the Aharonov-Bohm transformation that it generates
$$
U(C,x_0)=P\
{\rm exp}\biggl[ i\int _{(C, x_0)}dx^iA_i\biggr]  .
\eqn\flux$$
Here $x_0$ is an arbitrary basepoint
and $C$ is a loop starting and ending at $x_0$ that encircles
the string once in the counterclockwise direction
\refmark{ \preskill , \alford  ,\mba }.
The flux $U(C,x_0)$ must lie in the unbroken symmetry group
because the covariant derivative of the Higgs condensate
must vanish along the path $C.$ The Aharonov-Bohm interactions
of matter fields with the string are most easily
analysed using a basis in which $U(C,x_0)$ is diagonal,
so that $U$ acting on the matter fields takes the form
${\rm diag}[ e^{i\xi _1}, e^{i\xi _2},\ldots  ].$ Then the
scattering cross section for each component is given
by the classic result
$$
{d\sigma \over d\theta }={1\over 2\pi k}
{\sin ^2(\xi /2)\over \sin ^2(\theta /2)}.
\eqn\ab $$
When the incident beam is a superposition of diagonal components,
a gauge-dependent Aharonov-Bohm scattering amplitude
must be used, and the relative phases between the components
are relevant.

When the magnetic flux carried by the string $U(C,x_0)$ lies
in the center of the unbroken symmetry group $H(x_0)$
(which we shall denote by $Z[H(x_0)]),$
there is little more to be said. However, when
$U(C,x_0)\notin Z[H(x_0)],$ new physics arises.
In this case, the cosmic string solution, considered
from the point of view of classical field theory,
is no longer invariant under the action of $H.$
This fact has two important consequences: (1)
The classical string solution has internal zero modes, which
lead to a manifold of classical degenerate string solutions.
The effect of these zero modes is most easily analyzed by
considering a loop of string. \footnote\dagger{To simply
the discussion, we shall ignore the translational degrees
of freedom (i.e., the ability of the loop to oscillate) and
consider a static string loop of radius $R.$}
When the loop is quantized, this classical internal degeneracy
is lifted, and a spectrum of charged string loop states
emerges. This charge is ``Cheshire" charge, discussed
classically in refs.
\preskill ~ -\ginsparg ~
and quantum mechanically in refs. \bucher ~ and \mbb .
``Cheshire" charge is peculiar because it is nonlocalizable;
it does not reside on the string, nor can it be attributed
to a current source in the vicinity of the string. Rather,
it is a sourceless charge due to the peculiar
matching condition that arise owing to the magnetic flux carried
by the loop.  (2) The interaction of charged particles and the string
becomes more complicated and is no longer simply described
by equation \ab . Even for particles that travel in a narrow beam
near the string,
with no interference between paths of different winding
number around the loop, the loop distorts the electric
field of the particle, thus creating a new interaction.

The nature of Cheshire charge is most easily understood in
the context of $U(1)$ Alice strings. These exotic
strings arise in a model with
unbroken symmetry group $H=U(1)_Q\times Z_2,$
where the product is semidirect, and the generator
$X$ of $Z_2$ conjugates $U(1)_Q$ charge---that is,
$XQX^{-1}=-Q.$ Alice strings carry a magnetic flux that
lies in the
disconnected component $H_d=\{ Xe^{i\phi Q}\vert 0\le \phi <2\pi \} .$
Since under a continuous gauge rotation $\Omega =e^{i\xi Q},$
a magnetic flux $U=Xe^{i\phi Q}$ transforms into $\Omega U\Omega ^{-1}%
= e^{i\xi Q} [Xe^{i\phi Q}] e^{-i\xi Q}=Xe^{i(\phi -2\xi )Q},$
there is no preferred flux in $H_d,$ so there is a zero mode
leading to a manifold of degenerate classical solutions with
the topology of $S^1,$ which when quantized leads to a spectrum
of charged states. One may study the electrodynamics of a string
loop classically. Consider an electric charge $+q$ carried around a
closed path passing through a loop of Alice string. Its charge
seems to change from $+q$ to $-q,$ creating the appearance
that a charge  $+2q$ has somehow disappeared. The resolution
of the paradox is that the missing charge has been transferred to
the string loop in the form of nonlocalizable Cheshire charge.
At the classical level, Cheshire charge is possible because
of the twist in $U(1)_Q$ as one passes through the
loop. There exist vacuum solutions to Maxwell's equations
for which $(\nabla \cdot {\bf E})=0$ everywhere, while
at the same time the electric field
integrated over a surface surrounding the loop would lead one
to infer using Gauss's law that charge is enclosed by the surface.
Alice strings---at least for $U(1)$ identified
with ordinary electromagnetism---while interesting, are not very
realistic because in models with Alice strings
the $Z_2$ symmetry must be exact. One knows that
in the real world charge reflection is not an exact symmetry.
Therefore, the possibility of observing these strings can be ruled out.

However, $SU(3)$ color Cheshire charge, by contrast, does not require
any exotic new symmetries that must be reconciled with
experiment. In scenarios for grand unification, one
typically has a pattern of symmetry breaking
$$\tilde G\to \ldots \to U(1)_{EM}\times SU(3)_{color}\times D
=H_{cont}\times D.\eqn\pattern$$
Here $D$ is a discrete group.
To simplify the discussion of whether the theory has topologically-stable
cosmic string solutions, we take the group $\tilde G$ to
be a simply-connected universal covering group, even when $\tilde G$
does not act effectively on the fields of the theory. This choice of
$\tilde G$ has the consequence that the entire subgroup $D$ or
a subgroup thereof may act trivially on the matter fields of the
theory, thus avoiding the introduction of new {\it physical}
discrete symmetries to be reconciled with experiment.
Since $\pi _1[\tilde G/H]=D,$ to each nontrivial element of
$D$ there corresponds a topologically-stable cosmic string solution.
Without more details about the model,
one cannot determine precisely what magnetic
flux is carried by the string configuration of minimal
energy in the topological sector described by $d\in D.$
All that can be determined from the topology is that $U$
belongs to the coset $d[U(1)_{EM}\times SU(3)_{color}].$ We
may write $U_d=g_{U(1)}\otimes g_{SU(3)}\otimes d.$
If $g_{SU(3)}\notin Z_3,$ then the string solution carries colored
Cheshire charge.

In this paper, we discuss the quantization of strings with
non-Abelian Cheshire charge, and also discuss the process by
which Cheshire charge is transferred to string loops by
passing charged objects through the loop. Before proceeding
to a general discussion of non-Abelian Cheshire charge, we
discuss in section 2 an $SO(10)$ GUT model with cosmic string
solutions which support $SU(3)_{color}$ Cheshire charge. Although
the existence of cosmic strings in models of $SO(10)$ grand unification
had been discussed long ago, only recently was it demonstrated
by Ma that the magnetic flux that minimizes the string tension
in the first $Spin(10)\to SU(5)\times Z_2$ phase transition
takes a value that is not invariant under $SU(5).$\refmark{\ma }
Ma showed numerically that an ansatz with such an asymmetric flux leads
to a lower energy per unit length than an Abrikosov-Nielsen-Oleson
ansatz whose flux lies in the center of $SU(5)\times Z_2.$
In section 2 we calculate which flux orientation for these strings
is energetically preferred in the two subsequent phase transitions
$ SU(5)\times Z_2\to SU(3)\times SU(2)\times U(1)_Y\times Z_2\to
SU(3)\times U(1)_Q\times Z_2.$ We calculate the effect of the
flux on the fermions through an Aharonov-Bohm  scattering. Our conclusions
regarding Aharonov-Bohm scattering differ from those of the authors of
refs. \brand ~ and \ma . In particular, we note that
a long-distance Aharonov-Bohm effect which does not involve core penetration
cannot lead to processes forbidden by the unbroken symmetry group,
such as the $B$ to $L$ processes claimed to exist in
refs. \brand ~ and \ma . In section 3 we discuss colored
Cheshire charge. In section 4 we discuss vortex-vortex scattering
for the case where the unbroken symmetry group is discrete.
In section 5 we discuss Alice vortex-vortex scattering.

\vskip 15pt
\chapter{$SO(10)$ Strings and the Non-Abelian Aharonov-Bohm Effect}

A potentially realistic $SO(10)$ model of grand unification
can be constructed by the pattern of symmetry breaking
\def\mapright#1{\smash{\mathop{\longrightarrow}\limits^{#1}}}
$$
Spin(10)
\mapright{126}
SU(5)\times Z_2
\mapright{45}
SU(3)\times SU(2)\times U(1)_Y\times Z_2
\mapright{10}
SU(3)\times U(1)_Q\times Z_2,
\eqn\pat
$$
where
$$\eqalign{
\left<\phi _{126}\right> &= v_{126}\cdot[
(e_1+ie_2)\wedge (e_3+ie_4)\wedge (e_5+ie_6)\wedge
(e_7+ie_8)\wedge (e_9+ie_{10})] ,\cr
\left<\phi _{45}\right> &= v_{45}\cdot {\rm diag}
[ 2/3,~ 2/3,~ 2/3,~-1,~-1],\cr
\left<\phi _{10}\right> &= v_{10}\cdot [
0,~ 0,~ 0,~ 0,~ 0,~ 0,~ 1,~ 0,~ 0,~ 0~ ]. \cr
}\eqn\vevs$$
With this partial choice of gauge, the Cartan subalgebra
for $SU(3)\times SU(2)\times U(1)_Y$ in terms of the $SO(10)$
generators is
$$\eqalign{
T_3^{SU(3)}&={1\over 2}[M_{1,2}-M_{3,4}],\cr
T_8^{SU(3)}&={1\over 2\sqrt{3}}[-M_{1,2}-M_{3,4}+2M_{5,6}],\cr
T_3^{SU(2)_L}&={1\over 2}[M_{1,2}-M_{3,4}],\cr
Y&={2\over 3}[M_{1,2}+M_{3,4}+M_{5,6}]-
[M_{7,8}+M_{9,10}].\cr
}\eqn\cartan$$
The $SU(5)$ generators may be written as
$T^I_{ab}=[M_{(2a-1),(2b-1)}+M_{2a,2b}]$ for $a\le b$
$T^R_{ab}=[M_{2a,(2b-1)}-M_{2a-1,2b}]$ for $a<b,$
where $(a,b=1,\ldots ,5).$ A possible basis for those
$SO(10)$ generators not included in $SU(5)$ consists of the two
sets of ten generators each
$$\eqalign{T^A_{ab}&=[M_{(2a-1),(2b-1)}-M_{2a,2b}],\cr
T^B_{ab}&=[M_{2a,(2b-1)}+M_{2a-1,2b}]\cr }
\eqn\gen$$
where $a<b,$
and the lone generator
$T^{U}_{ab}=[M_{1,2}+M_{3,4}+M_{5,6}+M_{7,8}+M_{9,10}],$
which commutes with all elements of $SU(5).$

This GUT model has topologically-stable cosmic string
solutions
because of the $Z_2$ discrete symmetry generated by the $2\pi $
rotation in $SO(10),$ which we shall denote by $X.$
The element $X$ acts on
the matter fields of the theory in the following manner. All
fermions acquire a phase $-1$ under the action of $X$
because the fermions transform under a spinor representation
of $Spin(10),$ and $X$ acts trivially on bosons because all
bosons transform under tensor representations of $Spin(10).$
Therefore, in this model the $Z_2$ discrete symmetry does not
forbid or even constrain any processes not already forbidden by the
fermion/boson superselection rule. From topological arguments we know
that there exists a stable cosmic string solution carrying a
flux that lies in the coset $X[SU(5)].$ [For the moment,
we consider only the first stage of symmetry breaking
$Spin(10)\to SU(5)\times Z_2.$]
Energetic considerations determine which flux (or set of fluxes)
belonging to this coset minimizes the energy per unit length of the string.

To determine which flux is energetically preferred, one
must examine competing possibilities.
Aryal and Everett
considered the ansatz
$$\eqalign{
\phi _{126}(r, \theta )&=~e^{i\theta t^s}\
f(r)~\phi _{126}\cr
{\bf A}(r, \theta )&= \hat {\bf e}_\theta \ {g(r)\over r}\cdot t^s\cr
}\eqn\ansatzone$$
where $\phi _{126}$ is given in eqn. \vevs , and
$f(0)=g(0)=0,$ and $f(r),\ g(r)\to 1$ as $r\to \infty ,$
and $t^s=(1/5)T^U_{ab}.$ The generator
$t^s$ is certainly the most symmetric choice
since $X=e^{i2\pi t^s},$ which commutes with all elements of $SU(5).$
However, as pointed out by Ma, it is not the choice that
minimizes the energy. Ma considered the ansatz,
originally suggested by Aryal and Everett,\refmark{\aryal }
$$\eqalign{
\phi _{126}(r, \theta )&=e^{i\theta t^a}\
[ f^{ext}(r)~\phi _{126}^{ext}
+f^{int}(r)~\phi _{126}^{int}
]\cr
{\bf A}(r, \theta )&= \hat {\bf e}_\theta \ {g(r)\over r}\cdot t^a\cr
}\eqn\ansatztwo$$
where $t^a=(1/2)T^A_{1,2},$
$f^{ext}(0)=g(0)=0$ and $f^{ext}(r),\ g(r)\to 1$ and
$f^{int}(r)\to 0$ as $r\to \infty .$
$\phi _{126}^{ext} =\phi _{126}$
and
$$\phi _{126}^{int} = v_{126}\cdot
[(e_1\wedge e_3-e_2\wedge e_4)\wedge (e_5+ie_6)\wedge
(e_7+ie_8)\wedge (e_9+ie_{10})] .
$$
Ma found that for all potentials with at most quartic terms
this ansatz leads to a lower energy per unit length
than the ansatz \ansatzone .

{}From the point of view of magnetic energy, the flux direction
$t=t^s$ is preferred over $t=t^a$ because
$ {\rm tr}[(t^s)^2]< {\rm tr}[(t^a)^2].$
However, in ansatz \ansatztwo ~ the Higgs field does not vanish in the
core of the string, and
$\phi _{126}^{int}$
has the property that
$t^a\cdot \phi _{126}^{int}=0.$ Therefore, the magnetic field
in the core does not introduce any scalar gradient energy
for that part of the scalar field pointing in the
$\phi _{126}^{int}$ direction. This means that if for
some $\lambda \ne 0,$ $V[\lambda \phi _{126}^{int}]<
V[0]=0,$ then ansatz \ansatztwo ~ can lead to a
lower energy by having a larger core, which numerically
has been shown to be the case.

We now consider how classical field configurations described by
ans\"atze \ansatzone ~ and \ansatztwo ~
transform under the action of the unbroken symmetry
group $SU(5).$ A configuration described
by \ansatzone  ~ is invariant under the action of $SU(5),$
but the same is not true for \ansatztwo .
A global gauge transformation $\Omega \in SU(5)$
acts on $t^a$ according to the rule $t^a\to \Omega t^a\Omega ^{-1}.$
By starting with
$t^a=t^a_0=(1/2)T^A_{1,2}$ and acting on $t^a$ with global gauge
transformations, we can obtain the assignments
$t^a=(1/2)=T^A_{a,b}$
and
$t^a=(1/2)=T^B_{a,b}$ for arbitrary $a,$ $b,$ as well as other flux
assignments. If there were no subsequent stages of symmetry
breaking, all such solutions would have degenerate energy
at the classical level.

In the two subsequent stages of symmetry breaking---first
to $SU(3)\times SU(2)\times U(1)\times Z_2,$ and then to $SU(3)\times
U(1)\times Z_2$---the freedom to make global gauge transformation
becomes restricted, and different orientations of $t^a$
need not lead to solutions of degenerate energy,
because interactions with the fields $\phi _{45}$ and
$\phi _{10}$ can, and generically do, break the degeneracy.
Therefore, to determine the lowest energy solution, we
must first determine the preferred orientation of $t^a$
within the conjugacy class
$[t^a]=\{ \Omega t^a\Omega ^{-1}\vert \Omega \in SU(5)\} $
in light of the subsequent stages of symmetry breaking.

In general, a subsequent stage of symmetry breaking
$G_1\to G_2$ through the condensation of a field $\Phi $
can have two effects. Suppose that $U\in G_1$ and solutions
with the fluxes in the conjugacy class
$[U]=\{ \Omega U\Omega ^{-1}~\vert \Omega \in G_1 \} $
are degenerate when the condensation of the field $\Phi $
is not taken into account. If $[U]\cap G_2$ is empty, then all
flux orientations would frustrate the Higgs field $\Phi $
at large distances (which would lead to an infinite energy
per unit length), and consequently the cosmic string solution
acquires an outer core
with a $G_2$ screening flux to avoid large distance frustration
of the field $\Phi .$ Alternatively, suppose that the intersection
$[U]\cap G_2$ is nonempty. Then these flux orientations
are preferred because no
$G_2$ screening flux is required. However, inside the core the
field $\Phi $ may still be
frustrated where there is a magnetic field, and those
orientations that minimize this frustration are
energetically preferred.

\def\diag#1{{\rm diag}[#1]}

For the second stage of symmetry breaking, the three
flux orientations
$t^a_{(1)}=(1/2)T_{1,2}^A,$
$t^a_{(2)}=(1/2)T_{3,4}^A,$
and $t^a_{(3)}=(1/2)T_{4,5}^A$
are not related to each other by conjugation by
$SU(3)\times SU(2)\times U(1)\times Z_2.$
Although
$\exp [i2\pi t^a_{(k)}]\in SU(3)\times SU(2)\times U(1)\times
Z_2$ for $(k=1,2,3),$
$[t^a_{(k)},\phi _{45}]
=v_{45}~[t^a_{(k)},Y]\ne 0.$
The generator that minimizes
${\rm tr}
\left(
[t^a_{(k)},Y]^2
\right) $
is energetically preferred. A simple calculation shows
that of these three choices $t^a=(1/2)T^A_{34}$ gives
the energetically preferred orientation;
therefore, with respect to the second stage of symmetry
breaking, the orientations
$t_a=(1/2) \Omega ~T^A_{3,4}~\Omega ^{-1}$ where
$\Omega \in SU(3)\times SU(2)\times U(1)$ are energetically
preferred. Note that both
$t_a=(1/2)T^A_{3,4}$ and
$t_a=(1/2)T^A_{3,5}$
are included in this set.
We now consider the final stage of symmetry breaking.
The choice
$t_a=(1/2)T^A_{3,4}$
is unacceptable because
$\exp[i2\pi(1/2)T^A_{3,4}]\notin
SU(3)\times U(1)_Q\times Z_2 .$
However,
$\exp[i2\pi(1/2)T^A_{3,5}] =
\diag{1,1,-1,1,-1}_{10}\in
SU(3)\times U(1)_Q\times Z_2 .$
Finally, we note that for
$t^a=(1/2)T_{3,5},$ $\exp[i2\pi t_a]$
is not invariant under the action of $SU(3)_{color},$
and although $\exp[i2\pi t_a]$ is invariant under the
action of $U(1)_Q,$ $t_a$ is not invariant under the action of
$U(1)_Q.$ The consequences of this lack of invariance
of the classical solution under unbroken continuous symmetries
will be discussed in the next section. We now discuss the
Aharonov-Bohm interaction of fermions with
a classical string background.

A single fermion generation is introduced as an
irreducible multiplet of the 16-dimensional spinor
representation. With the following basis for the Clifford algebra
$$\eqalignno{
\Gamma _1&= \sigma _1 \otimes \sigma _3 \otimes
\sigma _3 \otimes \sigma _3 \otimes \sigma _3 ,\cr
\Gamma _2&= \sigma _2 \otimes \sigma _3 \otimes \sigma _3 \otimes
\sigma _3 \otimes \sigma _3 ,\cr
\Gamma _3&= 1\otimes \sigma _1 \otimes \sigma _3 \otimes
\sigma _3 \otimes \sigma _3 ,\cr
\Gamma _4&= 1\otimes \sigma _2 \otimes \sigma _3 \otimes
\sigma _3 \otimes \sigma _3 ,\cr
\Gamma _5&= 1\otimes 1\otimes \sigma _1 \otimes \sigma _3 \otimes
\sigma _3 ,\cr
\Gamma _6&= 1\otimes 1\otimes \sigma _2 \otimes \sigma _3 \otimes
\sigma _3 ,\cr
\Gamma _7&= 1\otimes 1\otimes 1\otimes \sigma _1 \otimes \sigma _3 ,\cr
\Gamma _8&= 1\otimes 1\otimes 1\otimes \sigma _2 \otimes \sigma _3 ,\cr
\Gamma _9&= 1\otimes 1\otimes 1\otimes 1\otimes \sigma _1,\cr
\Gamma _{10}&= 1\otimes 1\otimes 1\otimes 1\otimes \sigma _2 ,\cr }$$
the particle content becomes
$$\eqalignno{
\vert \nu ^c\rangle &=\vert +,~ +,~ +,~ +,~ +~\rangle \cr
\vert \nu \rangle &=\vert -,~ -,~ -,~ +,~ -~\rangle \cr
\vert e \rangle &=\vert -,~ -,~ -,~ -,~ +~\rangle \cr
\vert e^c \rangle &=\vert +,~ +,~ +,~ -,~ -~\rangle \cr
\vert u_1\rangle &=\vert +,~ -,~ +,~ +,~ -~\rangle \cr
\vert u_2\rangle &=\vert -,~ +,~ +,~ +,~ -~\rangle \cr
\vert u_3\rangle &=\vert +,~ +,~ -,~ +,~ -~\rangle \cr
\vert d_1\rangle &=\vert +,~ -,~ +,~ -,~ +~\rangle \cr
\vert d_2\rangle &=\vert -,~ +,~ +,~ -,~ +~\rangle \cr
\vert d_3\rangle &=\vert +,~ +,~ -,~ -,~ +~\rangle \cr
\vert u_1^c\rangle &=\vert -,~ +,~ -,~ +,~ +~\rangle \cr
\vert u_2^c\rangle &=\vert +,~ -,~ -,~ +,~ +~\rangle \cr
\vert u_3^c\rangle &=\vert -,~ -,~ +,~ +,~ +~\rangle \cr
\vert d_1^c\rangle &=\vert -,~ +,~ -,~ -,~ -~\rangle \cr
\vert d_2^c\rangle &=\vert +,~ -,~ -,~ -,~ -~\rangle \cr
\vert d_3^c\rangle &=\vert -,~ -,~ +,~ -,~ -~\rangle \cr }$$

Since $M_{ij}=(-i/4)[\Gamma _i,\Gamma _j],$
$$
\left( t^a \right) _{16}=
\left( {1\over 2}T^A_{3,5} \right) _{16}
=\left( {1\over 2}[M_{5,9}-M_{6,10}]\cdot
\right) _{16}={1\over 4}[
1\otimes 1\otimes \sigma _1 \otimes \sigma _3 \otimes \sigma _2-
1\otimes 1\otimes \sigma _2 \otimes \sigma _3 \otimes \sigma _1]
\eqn\zxz$$
Thus $t^a$ acts nontrivially on eight fermions
as follows:
$$\eqalign{
t^a\vert \nu ^c\rangle &=+(i/2)\vert u_3\rangle ,\cr
t^a\vert u_3\rangle &=-(i/2)\vert \nu ^c\rangle ,\cr
t^a\vert u^c_3\rangle &=+(i/2)\vert \nu \rangle ,\cr
t^a\vert \nu \rangle &=-(i/2)\vert u^c_3\rangle ,\cr
t^a\vert d^c_1\rangle &=+(i/2)\vert d_2\rangle ,\cr
t^a\vert d_2\rangle &=-(i/2)\vert d^c_1\rangle ,\cr
t^a\vert d^c_2\rangle &=+(i/2)\vert d_1\rangle ,\cr
t^a\vert d^1\rangle &=-(i/2)\vert d^c_2\rangle .\cr
}\eqn\fgen $$
The other eight fermions are annihilated by $t_a.$

{}From the relations in equation \fgen ~ it may
appear that the vector potential rotates certain
quarks into leptons and vice versa. In fact,
certain authors argue that strings of this sort can
transform quarks into leptons by means of a
long-range Aharonov-Bohm interaction that
is completely of a topological nature, without
any core penetration. \refmark{\brand,\ma}
However, this phenomenon does not occur
because, while the fermion multiplet
twists as one passes around the string, the
Higgs fields also twist. Consequently,
at the same time that the twisting described
in equation \fgen ~ takes place, the definition
of what is a quark and what is a lepton changes
too, so that the purported process does not take
place.

The crucial point is that away from the core
of the string (except for exponentially decaying corrections)
there is pure vacuum. The winding in the Higgs field and the
non-vanishing gauge field has no local physical effect because
the covariant derivative of the Higgs field and the curvature
in the gauge field vanish.  Away from the core of the
string
in a simply-connected region $R$
one can make a unitary choice of gauge
so that the
Higgs field is constant and the gauge field $A_\mu $
vanishes.  However, such a unitary choice of gauge is not
possible in a multiply-connected region that encloses the string
because the winding of the Higgs field (and also the vector potential)
around the string poses a topological obstruction to a choice of
gauge in which $\phi (x)$ is everywhere constant. It is this
topological obstruction that can give rise to an Aharonov-Bohm
interaction between the string and the matter fields of the
theory.

The Aharonov-Bohm interactions of a cosmic string are most easily
analyzed in a singular gauge, in which all of the winding in
the Higgs field and the non-vanishing gauge potential are
concentrated on a singular sheet of zero thickness whose boundary
coincides with the string. In this singular gauge, there are nontrivial
matching conditions for the fields of the theory on opposite
sides of the sheet, more specifically
$$
\Psi (r, \theta = 2\pi )=e^{i2\pi \hat t^a}\ \Psi (r, \theta =0).
\eqn\zg $$
Here $\Psi $ is a 16 multiplet.
This matching condition is derived by requiring that the
{\bf covariant} derivative of $\Psi $ be nonsingular on the
sheet. This requirement is reasonable because the position of the
sheet is a gauge artifact and therefore of no physical
significance.

In the models described above, and the model considered in reference
\brand ~  as well,
the generator
$t^a$ can be put into a block
diagonal form in which the diagonal consists exclusively
of zeros and $2\times 2$ matrices of the form
$$
t^a={1\over 2}
\left(
\matrix{
0&-i\cr
i&0\cr
}
\right) \eqn\za
$$
that act on the multiplets
$(\nu ^c, u_3),$
$(u^c_3 , \nu ),$
$(d_1^c, d_2),$ and
$(d_2^c, d_1).$
One obtains the matching
conditions
$$\eqalign{
\nu (r,\theta =2\pi ,z)&=-\nu (r,\theta =0,z)\cr
\nu ^c(r,\theta =2\pi ,z)&=-\nu ^c(r,\theta =0,z)\cr
u_3(r,\theta =2\pi ,z)&=-u_3(r,\theta =0,z)\cr
u_3^c(r,\theta =2\pi ,z)&=-u_3^c(r,\theta =0,z)\cr
d_1(r,\theta =2\pi ,z)&=-d_1(r,\theta =0,z)\cr
d_1^c(r,\theta =2\pi ,z)&=-d_1^c(r,\theta =0,z)\cr
d_2(r,\theta =2\pi ,z)&=-d_2(r,\theta =0,z)\cr
d_2^c(r,\theta =2\pi ,z)&=-d_2^c(r,\theta =0,z)\cr
}\eqn\zb $$
The other fields have trivial matching conditions.
These conditions imply that
the particles
$\nu ,$ $\nu ^c,$ $u_3,$ $u_3,$ $d_1,$ $d_1^c,$ $d_2,$ and $d_2^c$
scatter off the string with the cross section
$$
{d\sigma \over d\theta }=
{1\over 2\pi k}\cdot
{1\over \sin^2(\theta /2)},
\eqn\zc
$$
corresponding to an Aharonov-Bohm phase $\xi =\pi .$
However, there is no mixing. In terms of flux,
$U=U_{SU(3)}\otimes e^{i3\pi Q}\otimes X$ where
$U_{SU(3)}=\diag{-1,-1,+1}.$

We have shown that in the particular models considered in
references \brand ~ and \ma ~ there is no $B$-violation
of a long-range topological origin. Note that our analysis
does not apply to short-range $B$-violation that takes
place inside the core. For small momenta, core penetration
is limited to the lowest partial wave. However, fractional
angular momentum wave function enhancement can cause
the core penetration cross section to be larger than
the naive geometric cross section, actually growing as a fractional
power of the wavelength.\refmark{wzb }
We now turn to the question of whether it is possible to construct
other models in which there is a $B$-violating Aharonov-Bohm
cross section of a topological nature. One might contemplate
the possibility that even though the models in
references \brand ~ and \ma ~ do not have
Aharonov-Bohm $B$-violation, other realistic models
which do have it could be constructed. We show
that this is not possible if $U(1)_Q$ is an exact symmetry
which can be globally defined.

Since leptons (and antileptons) have electric charge
$q=-1,0, +1,$ and quarks (and antiquarks) have electric charge
$q=-2/3,-1/3,+1/3,+2/3,$ a process that changes a quark into a
lepton (or antilepton) cannot conserve electric charge. For
processes that take place near the core of the string the missing
charge
can be transferred to the string core, where $Q$ is not
necessarily a good symmetry. However, for Aharonov-Bohm scattering
the $S$ matrix must leave all of the Higgs fields invariant.  Since
a transition from quark to lepton obviously fails to commute with
the Higgs field which breaks $SU(5)$ to $SU(3)\times SU(2)\times
U(1),$ such a process must be forbidden.  In fact,
even before we study
the possibility that cosmic strings could be Alice strings, we already
know that the associated delocalized Cheshire charge could not be
ordinary electric charge:  There is no operator which both
changes electric charge and commutes with all the Higgs fields which
must be present in any theory containing the standard model.

\vskip 15pt
\chapter{Colored Alice Strings}

It is well known that when a topological defect solution---or
soliton---in classical field theory is not invariant under
the action of a continuous unbroken internal symmetry, there exist
zero modes that lead to a manifold of classically degenerate
solutions.  In the quantum theory the classical
degeneracy is broken, so that in addition to the $H$ invariant
soliton, there exist dyonic states that carry nontrivial $H$ charge.
The classic example, first discussed by Julia and Zee,
\refmark{\julia }
is the `t Hooft-Polyakov monopole, whose core is not invariant
under $U(1)_Q$ rotations. The result is the existence of
electrically-charged dyonic excitations, which carry
magnetic charge $g$ and electric charge $ne,$
where $n$ is an integer. Another example is the
Skyrmion\refmark{\sky }.

The same phenomenon occurs for cosmic strings. When the flux
$U\in H$ of a cosmic string does not lie in the center of
$H,$ loops of cosmic string can carry electric charge.
The classic example
is the Alice string, which can carry nonlocalizable Cheshire
charge \refmark{\preskill ,\alford ,\ginsparg }. For Alice strings
\refmark{\schwarz },
the unbroken symmetry group is $H=U(1)_Q\times Z_2,$
where the product is semidirect with the generator
$X$ of $Z_2$ conjugating $U(1)_Q$ charge---that is,
$XQX^{-1}=-Q.$ Alice strings carry a magnetic flux which
lies in the
disconnected component $H_d=\{ Xe^{i\phi Q}\vert 0\le \phi <2\pi \} .$
Since under a continuous gauge rotation $\Omega =e^{i\xi Q},$
a magnetic flux $U=Xe^{i\phi Q}$ transforms into $\Omega U\Omega ^{-1}%
= e^{i\xi Q} [Xe^{i\phi Q}] e^{-i\xi Q}=Xe^{i(\phi -2\xi )Q},$
there is no preferred flux in $H_d.$ At the classical
level, there is a degeneracy, with each flux $U\in H_d$
corresponding to a distinct classical cosmic string solution
of degenerate energy \refmark{\bucher }. Quantum mechanically,
this degeneracy is broken.
The energy eigenstates are states of definite charge
$2N$
$$\vert N> =\int _0^{2\pi }d\phi \ e^{i\phi N}\vert \phi >.\eqn\cc $$
Formally, quantizing the Alice string zero mode is precisely
analogous to quantizing a rotor in two dimensions.

This discussion of Cheshire charge so far has been rather abstract,
based on the transformation properties of magnetic flux. For Alice
strings the existence---and also necessity---of Cheshire charge
already is manifest at the level of classical field theory.
As mentioned in the introduction,
loops of string must be able to carry
nonlocalizable Cheshire charge in order to resolve a paradox in which charge
conservation would appear
to be violated. Consider a charge $+q$ carried around a
closed loop passing through a loop of Alice string. Its charge
appears to change from $+q$ to $-q,$ creating the appearance
that a charge $+2q$ has somehow disappeared. The resolution
of the paradox is that the missing charge has been transferred to
the string loop in the form of nonlocalizable Cheshire charge.
At the classical level, Cheshire charge is possible
because of the twist in $U(1)_Q$ as one passes through the
loop. There exist vacuum solutions to Maxwell's equations
for which $(\nabla \cdot {\bf E})=0$ everywhere, while
at the same time the electric field
integrated over a surface surrounding the loop would lead one
to infer using Gauss's law that there is charge enclosed.

Similar phenomena occur for the colored $SO(10)$
strings discussed in the previous section. The flux of the strings
discussed in the previous section is described by
a matrix ${U^a}_b,$ which transforms as
$U\to \Omega U\Omega ^{-1}$ under a gauge rotation
$\Omega \in SU(3).$ The matrix $U$ after an appropriate
gauge rotation has the form \footnote\dagger{This is what $U$
acting on the up quarks looks like. For the down quarks there
is an additional overall factor of $(-1)$, but this does not
change the discussion of colored Cheshire charge.}
$$U={\rm diag}[1,1,-1].\eqn\da $$
Physically, the orientation of $U$ in internal color
space is measurable; it indicates which color
of quark experiences nontrivial Aharonov-Bohm
scattering.

Classically, the cosmic string solution has
zero modes. Since $SU(3)$ acts nontrivially on the classical
string solution, there exists a manifold of classically
degenerate string solutions. The possible color orientations
described by $U$ lie in the coset space
$$M\cong {H\over H_{inv}}={SU(3)\over [SU(2)\times U(1)]/Z_2}.\eqn\zl$$
Here the subgroup $H_{inv}\subset H$ is the part of the unbroken
symmetry group that leaves $U$ invariant.

Quantizing the color zero mode is analogous to
the abelian Alice string, except that
the representations are slightly more complicated.
The Alice string loop zero mode was like that of a spinning top
in two dimensions.
The colored string loop is more like the symmetric
rigid rotor in three dimensions, except that
the symmetry group is $SU(3)$ instead of $SU(2),$
and also a quotient over a subgroup has been taken.

Quantum mechanically, the state of the flux degrees of
freedom of the string loop is described by a
wave function $\Psi (m),$ whose domain is the manifold of classically
degenerate vacua $M,$ defined in \zl . States
of definite $SU(3)$ charge---that
is, states which transform irreducibly under $SU(3)$---correspond
to functions $Y^a_{\bar mi}(m),$ where the index $\bar m$
labels the irreducible
representations of $SU(3).$ Let $d^{\bar m}$ be the dimensionality
of the representation $\bar m.$ The index $i$ ranges from $1$ to $d^{\bar m}.$
The index $a$ is included to account for multiple copies of
$\bar m.$ Square-integrable functions on $M$ can be expanded as
$$
\psi (m)=\sum _{\bar m, a, i}C^{\bar m}_{a,i}Y^{\bar m}_{ai}(m).\eqn\su $$

To determine what kinds of $SU(3)_{color}$ charge the loop may carry, it is
convenient to represent the function $\psi (m)$ as a function
whose domain is $SU(3)$ instead of the coset space $M.$ We consider
functions $\psi :SU(3)\to {\cal C}$ that are invariant under
translation by $H_{inv}=[SU(2)\times U(1)]/Z_2.$ The Peter-Weyl
theorem \refmark{\barut } implies that an arbitrary function
on $SU(3)$ may be expanded in terms of the complete orthonormal set
$\sqrt{d^{\bar m}}D^{\bar m}_{aa^\prime }(g).$ Let $N^{\bar m}$
be the number of times that the representation $\bar m$
occurs in the expansion \su .
$$\eqalign{N^{\bar m}\ \ &=\
\int _{g^\prime \in {SU(2)\times U(1)\over Z_2}}dg^\prime \
\int _{g\in SU(3)}dg\
D^{\bar m}_{aa^\prime }(g^\prime g))
[D^{\bar m}_{aa^\prime }(g)]^\dagger \cr
&=\ \int _{g^\prime \in {SU(2)\times U(1)\over Z_2}}dg^\prime \ \
D^{\bar m}_{aa^{\prime \prime }}(g^\prime )
\int _{g\in SU(3)}dg\
D^{\bar m}_{a^{\prime \prime }a^\prime }(g)
[D^{\bar m}_{aa^\prime }(g)]^\dagger \cr
&=\ \int _{g^\prime \in {SU(2)\times U(1)\over Z_2}}dg^\prime
\ \
\Gamma ^{\bar m}(g^\prime ).\cr }\eqn\zz$$
The integral over $g^\prime $ imposes invariance under
translations by $H_{inv}.$ Because of the orthonormality
relation for characters, the last line indicates that
$N^{\bar m}$ is simply the number of times
that the trivial representation
occurs in the reducible representation of $[SU(2)\times U(1)]/Z_2$
induced by the irreducible representation of $SU(3)$
labeled by
$\bar m.$ It is clear that representations of nontrivial
triality cannot occur because the center $Z_3$ of $SU(3)$
is included in $[SU(2)\times U(1)]/Z_2.$

Charge can be transferred to the loop by passing objects with
nontrivial $SU(3)_{color}$ quantum numbers through it. To see
this, consider an initially uncharged loop, described by the
wave function
$$
\vert \psi ^{initial}_{loop}>
{}~=\int _{m\in M} dm\ \vert m>,
\eqn\wa$$
and a color singlet quark-antiquark pair, described by the
wave function
$$
\vert \psi ^{initial}_{Q\bar Q}>~=
{1\over \sqrt{3}}\delta _i^j
\vert Q_j>
\vert \bar Q^i>,
\eqn\wb$$
so that the initial wave function for the color degrees of
freedom of the entire system is
$$
\vert \psi ^{initial}>=\vert \psi ^{initial}_{Q\bar Q}>
\otimes \vert \psi ^{initial}_{loop}>,
\eqn\wc$$
a simple tensor product. Initially, there are no correlations between the
two subsystems.

We now adiabatically transport the quark $Q$ through the
loop, while keeping the antiquark $\bar Q$ in place,
finally bringing the quark $Q$ back to its original
position, as indicated in fig. 8 \footnote\dagger{Actually,
it is unnecessary to carry out the process
adiabatically. As long as the quark propagates through
the loop and the antiquark $\bar Q$ does not, the wave
function discussed above describes the $SU(3)_{color}$
degrees of freedom exactly.}

The wave function now is
$$
\vert \psi ^{final}>
=\int _{m\in M} dm\
{1\over \sqrt{3}}U_i^j(m)\
\vert Q_j>
\vert \bar Q^i>\otimes
\vert m>,
\eqn\wd$$
The state
$\vert \psi ^{final}>$ is no longer a simple tensor product;
correlations have been established between the color degrees
of freedom of the loop and of the quark-antiquark pair. These
correlations are necessary to insure that the total color
charge of the entire system remains trivial, as it must
be owing to charge conservation.

Suppose that we now measure the $SU(3)_{color}$ charge of the
$Q\bar Q$ pair in the final state. We first calculate the
density matrix, by taking the trace over the string
loop degrees of freedom. The result is
$$
\hat \rho ^{final}_{Q\bar Q}={1\over 3}\int _{m\in M}
U^i_j(m) U^l_n(m)^\dagger \
\vert Q^j>
\vert \bar Q_i>
<\bar Q_l\vert
<Q^n\vert .\eqn\we $$
To calculate the probability $p^{final}_{singlet}$
of the final $Q\bar Q$ pair being
in a singlet state, we use the projection operator
$$
\hat P_{singlet}=
{1\over 3}
\vert Q^i>
\vert \bar Q_i>
<\bar Q_j\vert
<Q^j\vert \eqn\wf$$
to obtain
$$\eqalign{
p^{final}_{singlet}&=
{\rm tr}[\hat P_{singlet}\hat \rho ^{final}_{Q\bar Q}]={1\over 9},\cr
p^{final}_{octet}&=1-p^{final}_{singlet}={8\over 9}.\cr }\eqn\wg $$
These probabilities apply to the loop as well, because charge
conservation mandates that the combined system remain in a
color singlet state.
The generalization to more complicated situations where
the loop and the pair are initially charged or where objects
with color charge belonging to a different representation are passed
through the loop is straightforward.

So far we have ignored the effects of confinement and treated
$SU(3)_{color}$ almost as if it were a global symmetry.
This approximation is justified for a large range of length
scales, because the string thickness is of order $[10^{16}GeV]^{-1},$
while the effects of confinement become relevant for length scales
greater than
approximately $\Lambda _{QCD}^{-1}\approx [10^{-1}GeV]^{-1}.$
Of course, for loops larger than $\Lambda _{QCD}^{-1},$
confinement is relevant.
On scales much large than $\Lambda _{QCD}^{-1},$ there are
no colored objects; particles are color singlets. However,
the possibility of diffractive scattering of hadrons by the
string, with cross sections of order $\Lambda _{QCD}^{-1},$
exists. Consider a hadron that propagates very close to the
string in such a way that some of the partons pass around one side
of the string while the rest of the partons pass around the other side.
After this happens, a hadron, which previously was a color singlet, ceases to
remain a color singlet. Cheshire charge has been transferred from the hadron
to the string loop, but since confinement does not allow isolated
color charge, either the hadron bounces back elastically or
becomes excited to form a resonance, or if there
is enough energy a string of electric color flux forms, which
eventually fragments to form a jet of hadrons. Since cosmic strings
in an astrophysical context typically travel at a fraction of the
speed of light, and near cusps very close to the speed of light,
such interactions are plausible.

In the above we have
considered the transformation properties of
the magnetic flux
$U=\exp [i2\pi t^a]$
under
$SU(3)\times U(1),$
but we have not considered the transformation
properties of Lie algebra element $t^a$ generating
$U$ under $SU(3)\times U(1).$
These transformation properties are not necessarily the
same because for a given $U$ there may be several
choices of $t$ that give $U=\exp [i2\pi t].$
For example, for the $Spin(10)$ strings presented in
section 2,
$
\exp \left[ i(2\pi ){1\over 2}
[\cos \xi ~T^A_{3,5}+\sin \xi ~T^B_{3,5}]\right] $
is independent of $\xi ,$
so there is a continuum of such choices with the
topology of $S^1.$
Consider the action of $Q$ on $T^A_{3,5}.$ Recall
that since
$Q_{10}=\diag {+1/3,+1/3,+1/3,0,-1}$
and
$U_{10}=\diag {+1,+1,-1,+1,-1},$
the flux $U$ as seen away from the core of the string
is invariant under $U(1)_Q.$ However,
$[Q,T^A_{3,5}]={2\over 3}T_B^{3,5},$
and similarly
$[Q,T^B_{3,5}]=-{2\over 3}T_A^{3,5}.$
Therefore, the fields excited inside the string
core---the vector field, and the scalar field too---are
not invariant under the action of $U(1)_Q.$ This
fact implies that the strings are superconducting. The $SO(10)$
strings considered here exhibit both a `traditional' charged scalar
condensate in the core \refmark{\guta }
and a charge carrying condensate of vector
fields in the core\refmark{\everett }. Here both fields
are charged, because the direction $\phi _{126}^{int}$
also is not invariant under $U(1)_Q.$

One peculiar feature of these superconducting
strings is the quantization condition for the electric
charge on a string loop. The commutation
relations $[Q,T^A_{3,5}]={2\over 3}T^B_{3,5}$ and
$[Q,T^B_{3,5}]=-{2\over 3}T^A_{3,5}$ imply that
the charge $q$ on a loop satisfies $q={2\over 3}en$
where $n$ is an integer and $e$ is the magnitude
of the charge of an electron.

Because of the monopoles potentially present in the theory,
a loop state with fractional electromagnetic charge
and but trivial triality would violate the
Dirac quantization condition. Recall that our loop
states always carry trivial triality. The resolution
of this paradox is the following. In our discussion
of Cheshire charge we considered only the transformation
properties of the long range flux under $SU(3).$
We did not consider the transformation properties
of the generator $t^a$ under $SU(3).$ It turns out
that the subgroup $[SU(2)\times U(1)]/Z_2\subset SU(3)$
that leaves $U=\exp [i2\pi t^a]$ invariant does not leave
$t^a$ invariant. This feature allows loop states with fractional
electromagnetic charge to carry color charge corresponding to nontrivial
triality localized on the string itself,
so that the Dirac quantization condition is not violated.
It should be stressed that the electric charge and
the color charge of nontrivial triality is not Cheshire
charge, because it is localized on the string core. In
contrast to Cheshire charge, it is not sourceless; there
is a measurable, localizable charge on the string.

\vskip 15pt
\chapter{Non-Abelian Vortex-Vortex Scattering}

In its classic form, the Aharonov-Bohm effect involves
the scattering of charged particles by localized magnetic
flux where the charged particle is forbidden to
penetrate the region where the gauge field has curvature
(that is, where $F_{\mu \nu }\ne 0$), or where such penetration
can be neglected. For Abelian gauge fields, magnetic flux
is uncharged; however, for non-Abelian gauge fields,
this generally is not the case, since in general
$\Omega F_{\mu \nu }\Omega ^{-1}\ne F_{\mu \nu }.$  Thus it becomes
possible to consider Aharonov-Bohm experiments in which two straight,
parallel flux tubes pass by each other, moving only in the plane
normal to the tube axes.  Because this is such a strong restriction
on the allowed tube motions, the resulting
problem of vortex motion in two space dimensions may be considered artificial,
but perhaps still interesting because a great deal can be found about its
solution.

The possibility for nontrivial Aharonov-Bohm scattering
of non-Abelian magnetic flux by other non-Abelian
magnetic flux pointing in a different direction in internal
symmetry space is best illustrated by considering
vortex-vortex scattering in a theory where a gauge
symmetry $G$ is broken to a non-Abelian discrete subgroup $H$
by the formation of a Higgs condensate
\refmark{\wilczekwu - \preskilltwo }.
We shall take $G$ to be the simply-connected
covering group, so that to each nonunit element of $H$ there
corresponds a topologically stable vortex excitation.
Using a discrete little group $H$ alleviates certain complications,
which shall be discussed later.

Before considering the quantum theory, it is useful to
consider the classical description of a system with
such vortices. It is the manner in which the magnetic
flux carried by the vortices evolves classically as the vortices
are moved that gives rise
to nontrivial Aharonov-Bohm scattering in the quantized system.
A classical state of a system with $N$ vortices
at a time $t$ is specified by giving the positions
of the vortices ${\bf x}_1, {\bf x}_2,\ldots , {\bf x}_N$
and their magnetic fluxes as well. If the unbroken group $H$ is Abelian,
the magnetic fluxes can be specified by enclosing each
vortex once in the counterclockwise direction by a set of
closed paths $C_1,C_2,\ldots ,C_N$
and evaluating the fluxes
$$ h_j=P~~ \exp \biggl[ i\oint _{C_j}d{\bf x}\cdot {\bf A}({\bf x})\biggr]
\in H.  \eqn\flux$$
Here $P$ indicates that the exponential is path ordered. The flux
$h_j$ is an element of the unbroken symmetry group
and indicates the phase that results from adiabatically transporting
a particle around the vortex.

When $H$ is non-Abelian, a more elaborate formalism
is required because definition \flux ~ becomes ambiguous.
$h_j$ is no longer well defined because it too changes
upon being adiabatically transported around another vortex.
Upon parallel transport around a vortex with flux $h_i,$
one has $h_j\to h_ih_j{h_i}^{-1}.$ To avoid this
ambiguity, one must use paths that start and end at a
particular basepoint $x_0,$
as indicated in fig. 1. Expression \flux ~ is modified
to become
$$ h(C,x_0)=P~~ \exp \biggl[ ~i\int _{(C,{\bf x}_0)}d{\bf x}\cdot {\bf A}%
({\bf x})\biggr]
\in H({\bf x}_0).  \eqn\fluxtwo$$
Because the curvature of the gauge field vanishes everywhere except
at the vortex cores, the flux $h(C,x_0)$
is invariant under continuous deformations
of $C$ that avoid the vortex cores. Therefore, the expression
\fluxtwo ~ can be thought of as defining a mapping from the
fundmental group of the punctured plane
\def\x{{\bf x}}
${\cal M}={\cal R}-\{ \x _1,\x _2, \ldots ,\x _N\} $ to the little
group $H(\x _0)$ based at $\x _0.$
This mapping is a group homomorphism.
As shown in fig. 2, it is apparent that in the non-Abelian case
not only the winding number of a path around a vortex is relevant
in determining the flux $h(C,\x _0),$
but also its threading around the neighboring vortices. Paths
$\alpha $ and $\alpha ^\prime $ both have unit winding number around
the vortex $A$ and vanishing winding number around $B;$ however,
they differ in how they thread around the vortex $B.$
Since $\alpha ^\prime =\beta \alpha \beta ^{-1},$
$h( \alpha ^\prime )=h(\beta )h(\alpha )h(\beta )^{-1}.$
Therefore, $h( \alpha ^\prime )=h(\alpha )$ if and only
if the fluxes $h( \alpha )$ and $h(\beta )$ commute.

The formalism just outlined allows one to describe the magnetic fluxes
of a set of vortices at fixed time. A classical vortex
configuration at time $t_a$ is completely specified by the
positions of the vortices
$ {\bf x}_1(t_a), {\bf x}_2(t_a), \ldots ,{\bf x}_N(t_a)$
and the mapping
$h:\pi _1 [{\cal M}(t_a),\x _0]\to H(\x _0)$
where
${\cal M}(t_a)={\cal R}^2-\{
{\bf x}_1(t_a), {\bf x}_2(t_a), \ldots ,{\bf x}_N(t_a)\} $
We now consider what happens when the vortices move.
At a later time $t_b$ the state of the system is described
by their new positions
$ {\bf x}_1(t_b), {\bf x}_2(t_b), \ldots ,{\bf x}_N(t_b)$
and a new mapping
$h_{t_b}:\pi _1 [{\cal M}(t_b),\x _0]\to H(\x _0).$
\footnote\dagger{For simplicity we shall treat the vortices as
point particles that never overlap.} The new mapping is
completely determined by the trajectory
${\bf X}(t)=( {\bf x}_1(t), {\bf x}_2(t),\ldots , {\bf x}_N(t)~).$
The mapping $h_{t_b}$ is obtained from the mapping $h_{t_a}$
by taking paths in
$\pi _1[{\cal M}(t_b),\x _0]$
and dragging them back to
$\pi _1[{\cal M}(t_a),\x _0]$
using the trajectory ${\bf X}(t)$ to deform ${\cal M}(t).$

Now that we have succeeded in describing classical vortex
evolution, we turn to describing the quantized system.
The classical states introduced
above serve as a basis of orthogonal states for the Hilbert
space of physical states, analogous to the position eigenstates
$\vert x>$ in ordinary non-relativistic quantum mechanics.
To calculate the matrix element
$$
\left< ~
{\bf x}_1(t_b), {\bf x}_2(t_b), \ldots , {\bf x}_N(t_b);~h_{t_b}
{}~\vert ~
{\bf x}_1(t_a), {\bf x}_2(t_a), \ldots , {\bf x}_N(t_a);~h_{t_a}
{}~ \right>
\eqn\mat$$
for the $N$-vortex sector using the path integral formalism, we
sum over paths
${\bf X}(t)=( {\bf x}_1(t), {\bf x}_2(t),\ldots , {\bf x}_N(t)~)$
that in addition to the usual requirement that
${\bf X}(t)={\bf X}_a(t_a)$ for the initial state and
${\bf X}(t_b)={\bf X}_b(t_b)$ for the final state also satisfy the
requirement that ${\bf X}(t)$ causes $h_{t_a}$ to deform into $h_{t_b}.$
Paths satisfying the first requirement can be classified into
braid classes---that is, homotopy classes. The completely-colored
braid group describes the relation between different such
homotopy classes.
The last requirement restricts the sum to a subset of the
possible braid classes---namely those which give the proper
final flux $h_{t_b}.$
For a system with two vortices, $A$ and $B$ which
initially at $t$ are at positions $\x _a$ and $\x _b$
and propagate to positions $\x _a^\prime $ and $\x _b^\prime $
at a later time $t',$ paths ${\bf X}(t)$ connecting the initial
and final states may be classified
by a winding number $\bar N.$ Unless the initial and final positions
coincide, the choice of homotopy class corresponding to $\bar N=0$
is arbitrary. Increasing $\bar N$ by one unit involves allowing
vortex $A$ to wind around vortex $B$ by one more unit.

Vortex-vortex scattering for $N=2$ is illustrated in fig. 3.
For simplicity vortex $B$ is held fixed. As indicated in
fig. 3(b), vortex $A$ can propagate quantum mechanically
to its new position $A^\prime $ via many homotopically inequivalent paths,
two of which are indicated in the figure. Figs. 3(a) and 3(c)
indicate alternative flux bases for the initial state, and fig.
3(d) indicates a flux basis for the final state.
If vortex $A$ propagates from
$A$ to $A^\prime $ along path 1, then
$h(\alpha _f)=h(\alpha _1)$ and
$h(\beta _f)=h(\beta ).$
Alternatively, if vortex $A$ propagates from
$A$ to $A^\prime $ along path 2, then
$h(\alpha _f)=h(\alpha _2)$ and
$h(\beta _f)=h(\alpha \beta \alpha ^{-1}).$
Therefore, it is clear that the amplitudes do not add,
for they correspond to classically distinguishable final states.
Paths 1 and 2 are only two of infinitely many
homotopically inequivalent paths. To consider
other homotopy classes, let us consider the effect of increasing
the relative winding number $\bar N$ by one unit,
by the process indicated in fig. 4(b),
which we shall call ${\cal R}^2$ for reasons that shall become
apparent later.  One has
$$\eqalign{
\alpha _{(n+1)}&=
[\beta _{(n)}\alpha _{(n)}]
\alpha _{(n)}
[\beta _{(n)}\alpha _{(n)}]^{-1}=
[\beta _{(0)}\alpha _{(0)}]
\alpha _{(n)}
[\beta _{(0)}\alpha _{(0)}]^{-1}, \cr
\beta _{(n+1)}&=
[\beta _{(n)}\alpha _{(n)}]
\beta _{(n)}
[\beta _{(n)}\alpha _{(n)}]^{-1}=
[\beta _{(0)}\alpha _{(0)}]
\beta _{(n)}
[\beta _{(0)}\alpha _{(0)}]^{-1}. \cr }
\eqn\ft$$
The quantity $\beta _{(j)}\alpha _{(j)}$ is independent of $j$
because it is the path enclosing both vortices,
and the combined flux is a conserved quantity.
There exists a smallest $M \ge 1$ so that
$h(\alpha _{(M)})=h(\alpha _{(0)})$ and
$h(\beta _{(M)})=h(\beta _{(0)})$---in other words,
$M$ indicates the number of windings required to restore the
system to its original state.

We have shown that the classical configuration space of the
system may be thought of as an $M$-sheeted Riemannian surface
${\cal R}^2(M),$ described in polar coordinates by the metric
$ds^2=dr^2+r^2d\theta ^2$
where $\vert \theta \vert \le M\pi .$
There is a natural $M$-to-$1$ projection
$\pi : {\cal R}^2(M)\to {\cal R}^2$ into physical space.
For $m\in {\cal R}^2,$ the $M$ points
$\pi ^{-1}(m)$ in ${\cal R}^2(M)$ correspond to the $M$ values
of the magnetic flux that are possible at each possible
position of the vortex $A.$ The multiply-sheeted formalism
is a natural way to keep track of magnetic flux without
introducing unphysical cuts that are merely gauge artifacts.
Solving vortex-vortex scattering in the multiply-sheeted formalism
reduces to solving the Schroedinger equation for a free
particle in the multiply-sheeted spacetime, a mathematical
problem that was solved long ago in a different context,
that of rigorous diffraction theory. A. Sommerfeld and H. Carslaw
\refmark{\sommerfeld, \carslaw, \bornandwolf }
solved the Helmholtz equation $[\nabla ^2+k^2]\psi =0$
for an incident plane wave in an attempt to understand
rigorously diffraction by a semi-infinite absorbing screen.

To define `scattering,' one must first specify what
kind of propagation would constitute the absence of
scattering. For $M=1$ this is completely obvious; a plane
wave signifies the absence of scattering. For $M>1$ the plane
wave
$$
\psi _{pw}(r,\phi )=
\cases{
e^{-ikr\cos \phi },& for $\vert \phi \vert <\pi $,\cr
0,&for $\pi <\vert \phi \vert <M\pi $\cr }
\eqn\pw$$
does not satisfy the equations of motion, even in the absence
of a nontrivial potential. This is because in the multiply-sheeted space
the plane wave \pw ~ has sharp edges at $\phi =\pm \pi ,$ which
would correspond to an infinitely sharp shadow, as predicted by
geometric optics but contrary to the nature of wave propagation.
We define scattering by the asymptotic form
$$
\psi (r,\phi )=\psi _{pw}(r,\phi )+
{f(\phi )e^{ikr}\over \sqrt{r}},
\eqn\af$$
so that $\sigma (\phi )=\vert f(\phi )\vert ^2.$
Contrary to the usual convention, here $\phi =0$ points in
the backward direction, while $\phi =\pm \pi $ points in the
forward direction. This convention allows
one to distinguish between forward propagation to the left the scattering
center and forward propagation to the right of the scattering center.
For fixed $r$---irrespective of how large $r$ is---the asymptotic form \af ~
breaks down for $\vert \phi \vert $ sufficiently close to $\pi ,$
because the asymptotic form \af ~ represents ``far-field,"
or Fraunhofer,
diffraction, and therefore is valid only for
$\vert ~\vert \phi \vert -\pi ~\vert \gtorder \sqrt{(\lambda /r)}.$

Following Carslaw, we calculate $\psi (r,\phi ;M)$ exactly
\refmark{\carslaw }. We
re-express the plane wave (i.e., the $M=1$ solution) with the
contour integral
$$\psi (r,\phi -\phi ';M=1)=
e^{-ikr\cos [\phi -\phi ']}={1\over 2\pi }\int _C
{d\alpha \ e^{i\alpha } \over e^{i\alpha }-e^{i\phi }}\
e^{-ikr\cos [\alpha -\phi ']}\eqn\contone$$
where the contour
$C=C_{top}+C_{bottom}$
is sketched in fig. 5. The shaded regions indicate where the
integrand is vanishes as one approaches infinity. Note that
for equation \contone ~ the contributions from $C_{left}$ and $C_{right}$
cancel because of periodicity. (Later this will no longer be true.)
The equality of the
contour integral and the plane wave is demonstrated by adding
the contours
$C_{left}$ and $C_{right}$ to the original contour
$C,$ so that the contour becomes
closed and Cauchy's theorem can be applied.

We now modify \contone ~ by altering to integrand to change the
periodicity.
Equation \contone ~ is modified to become
$$\psi (r,\phi -\phi ';M)
={1\over 2\pi M}\int _C
{d\alpha \ e^{i(\alpha /M)} \over e^{i(\alpha /M)}-e^{i(\phi /M)}}
\ e^{-ikr\cos [\alpha -\phi ']}.\eqn\conttwo$$
For $M=\infty ,$ we define
$$\psi (r,\phi -\phi ';\infty )
={1\over 2\pi i}\int _C
{d\alpha \ \over (\alpha -\phi )}
\ e^{-ikr\cos [\alpha -\phi ']}.\eqn\contthree$$
Deforming $M$ away from unity spreads
out the simple poles of unit strength in the integrand, thus changing the
periodicity of $\psi ,$ so that
$\psi (r,\phi +2\pi M;M)=\psi (r,\phi ;M).$
\footnote\dagger{None of the arguments here
rely on $M$ being an integer. The situation where $M$ is not an
integer is of interest for studying wave propagation in the presence
of conical defects, which arise from point sources in $2+1$ dimensional
gravity and in the vicinity of cosmic strings in $3+1$ dimensional
gravity.} The variable $\phi ',$
which hereafter shall be set to zero,
allows one to show that the Helmholtz equation
is satisfied. Since $\psi $
depends only on the difference $(\phi -\phi '),$
$\partial _\phi ^2\psi =\partial _{\phi '}^2\psi ,$
so consequently $[\nabla ^2+k^2]\psi =0.$
It now only remains to be shown that
$\psi (r,\phi ;M)$ has the required asymptotic form.

The asymptotic form for equation \conttwo , defined
in equation \af , is derived using
the saddle point approximation, which
to leading order in $(kr)$ gives
$$
\int d\alpha ~F(\alpha )\exp \bigl[-(kr)G(\alpha )\bigr] =
\sqrt{\pi \over kr}
\sum _l
{(\pm )_l\over \sqrt{ G^{\prime \prime }(\alpha _l)}}
F(\alpha _l)
\exp \bigr[ -(kr)G(\alpha _l)\bigr]
\eqn\sp $$
where the index $l$ labels the relevant saddle points,
which for this problem, with $G(\alpha )=i\cos \alpha ,$
are located at $\alpha  =-\pi ,~0,~+\pi .$
As indicated in fig. 6(a), we deform the
contours in fig. 5 to pass through the saddle points.
In doing so, for $\vert \phi \vert <\pi $
it is necessary to push either $C_{upper}$ or $C_{lower}$ to
over a pole on the real axis in the interval $-\pi <\alpha <+\pi .$
The deformation around the pole adds an additional
closed contour, shown in fig.
6(b), giving a contribution equal to $\psi _{pw}$---which is
precisely the ``unscattered" wave. We now calculate the
contributions from the saddle point. The contributions
from $\alpha =0$ cancel, leaving
$$
\eqalign{
& {1\over 2\pi M}
\left( {1-i\over \sqrt{2}}\right)
\sqrt{2\pi \over kr}
e^{ikr}
\left[
{ e^{i(\pi /M)}\over e^{i(\pi /M)}-e^{i(\phi /M)}}-
{ e^{-i(\pi /M)}\over e^{-i(\pi /M)}-e^{i(\phi /M)}}
\right]
\cr
&=
\left( {1+i\over \sqrt{2}}\right)
{1\over \sqrt{2\pi kr}M}
e^{ikr}
\left[
{i ~\sin (\pi /M)\over
\cos (\pi /M) - \cos (\phi /M)} \right] ,\cr }
$$
so that the scattering amplitude in equation \af ~ is
$$
f_M(\phi )
=
\left( {1+i\over \sqrt{2}}\right)
{1\over \sqrt{2\pi k}M}
\cdot \left[
{\sin (\pi /M)\over
\cos (\pi /M) - \cos (\phi /M) }\right] .
\eqn\sa$$
The infinitely-sheeted scattering amplitude has the particularly
simple form
$$
f_\infty (\phi )
=
\left( {1+i\over \sqrt{2}}\right)
\sqrt{2\pi \over k}
\cdot
{1\over \phi ^2-\pi ^2 }.
\eqn\iassa $$
Equation \iassa ~ has also been derived by direct summation
of the Feynman path integral in polar coordinates with
the summation restricted to a particular homotopy class.
\refmark{\peak }
Equation \sa ~ is the amplitude for vortex-vortex scattering
where the initial and final states are both flux eigenstates.
Here $M$ indicates the minimal number of windings to recover
the original fluxes. As expected, the scattering amplitudes
\sa ~ and \iassa ~ have two forward peaks
at $\phi =\pm \pi $
where the cross section diverges, and away from these peaks
the amplitude decreases.
When $\vert \phi \vert \to \pi ,$ a pole approaches one of the
two relevant saddle points, eventually invalidating the saddle point
approximation made above for fixed $kr.$ This happens because
for $\vert \phi \vert $ close to $\pi $
at fixed $r$ sufficiently close to the
forward direction one sees a Fresnel diffraction pattern rather
than a small-angle divergence in the wave function.

It is interesting to note that with the infinitely-sheeted scattering
amplitude one can calculate the usual Aharonov-Bohm
scattering amplitude in a very simple way. Consider Aharonov-Bohm
scattering
of a particle of charge $q$ by an infinitely thin flux tube
carrying a flux $\Phi =(\alpha /2\pi )\Phi _0$ where $\Phi _0=%
(hc/q)$ is the quantum of flux with respect to the charge $q.$
It follows that
$$\eqalign{
f_{AB}(\phi ;\alpha )=&
\sum _{n=-\infty }^{+\infty }
e^{in\alpha }\ f_\infty (\phi +2\pi n)\cr
&=
\left(
{1+i\over \sqrt{2}}
\right)
\sqrt{2\pi \over k}
\sum _{n=-\infty }^{+\infty }
e^{in\alpha }\cdot
{1\over (\phi +2\pi n)^2-\pi ^2}.\cr }
\eqn\abamp$$
In other words, the Aharonov-Bohm scattering amplitude
is the superposition of contributions with all possible
winding numbers, added together with a relative phase
factor of $e^{i\alpha }$ for $\Delta N=+1.$
This sum is evaluated by considering contour integral
$$
{1\over 2\pi }\int _C {dz~
e^{i(\alpha /2\pi )z} \over e^{iz}-1}
\ {1\over (z+\phi )^2-\pi ^2}
\eqn\contint $$
where $C$ is a closed curve enclosing the entire real axis.
For $0 <\alpha <2\pi ,$ the integral \contint ~ vanishes,
so that the sum of the residues vanishes. In other words,
$$ \sum _{n=-\infty }^{+\infty }
e^{in\alpha }\cdot
{1\over (\phi +2\pi n)^2-\pi ^2}+
{1\over 2\pi }{\sin (\alpha /2)\over \cos (\phi /2)}e^{-i[(\alpha /2\pi )
-1]\phi }
=0.
\eqn\abresult$$
Because of periodicity, the result extends to all values
of $\alpha .$ Therefore we get
$$
f_{AB}(\phi ;\alpha )=
\left( {1+i\over \sqrt{2}} \right) {1\over \sqrt{2\pi k}}
{\sin (\alpha /2)\over \cos (\phi /2)}e^{-i[(\alpha /2\pi )
-1]\phi },
\eqn\finalamp$$
the classic Aharonov-Bohm result.
[Because of our peculiar convention for defining the scattering angle,
there is a cosine rather than a sine in the denominator.]
So far in our discussion of vortex-vortex scattering we have not
included the possibility of ``exchange" effects, whose relevance
was first pointed out by Lo and Preskill\refmark{\loandpreskill }.
One might not expect
exchange effects to be relevant for the non-Abelian vortex-vortex
interaction discussed here because vortices carrying the same
flux do not interact---at least not by means of the purely topological
effects we consider here. However, an exchange interaction does
sometimes arise for vortices carrying different fluxes.
In our discussion of two-vortex
scattering we considered only paths that take $A$ to $A^\prime $
and $B$ to $B^\prime .$ We did not consider paths that take
$A$ to $B^\prime $ and $B$ to $A^\prime .$ However, when $h(\alpha )$
and $h(\beta )$ lie in the same conjugacy class,
it is possible that amplitudes from the two types of paths must
be added together.

To consider this possibility, it is useful to consider the
braid operation ${\cal R},$ whose square was considered in
fig. 4. ${\cal R}$ exchanges two vortices in a counter-clockwise
sense, as indicated in fig. 7(a). With the flux basis indicated
in fig. 7(b), one has
$$
\eqalign{
{\cal R}&:\alpha \to \beta ,\cr
{\cal R}&:\beta \to \beta \alpha \beta ^{-1}.\cr
}\eqn\braid$$
Let us write
$\alpha _{(n+{1\over 2})}= {\cal R} \alpha _{(n)}$
and
$\beta _{(n+{1\over 2})}= {\cal R} \beta _{(n)}.$
Then, as before, define $\bar M$ to be the smallest
integer or half-integer such that
$\alpha _{(\bar M)}=\alpha _{(0)}$ and
$\beta _{(\bar M)}=\beta _{(0)}.$
If $M$ is a half-integer, then exchange effects are
relevant.  As a simple example, consider the symmetric
group $S^3.$ Suppose that $\alpha _{(0)}=(12)$ and
$\beta _{(0)}=(23).$ A simple calculation shows that
$\bar M=3/2.$ In the multiply-sheeted formalism, the
exchange contribution is taken into account very simply.
One just inserts the proper half-integral value of $M$
into the amplitude $\sa .$

\vskip 15 pt
\chapter{Abelian Alice Vortex-Vortex Scattering}

Let us consider vortex-vortex scattering of Abelian Alice strings,
for the moment ignoring the possibility of Cheshire charge,
which gives rise to a classical force, in addition to the
purely quantum-mechanical flux holonomy effect.

We start with two Alice vortices in flux eigenstates, so that
$$\eqalign{
&h(\alpha _{(0)})=X,\cr
&h(\beta _{(0)})=Xe^{i\xi Q}.\cr
}\eqn\aa$$
We have set the phase of
$h(\alpha _{(0)})$
to zero by an overall global gauge rotation. Recall
$$\eqalign{
&\alpha _{(j+{1\over 2})}=\alpha _{(j)}
\beta _{(j)} \alpha _{(j)}^{-1},\cr
&\beta _{(j+{1\over 2})}
=\alpha _{(j)},\cr
}\eqn\ab$$
so that one has
$$\eqalign{
&\hbox to 2in {$h(\alpha _{(0)})=X,$\hfill }
\hbox to 2in {$h(\beta _{(0)})=Xe^{i\xi Q},$\hfill }\cr
&\hbox to 2in {$h(\alpha _{({1\over 2})})=Xe^{-i\xi Q}
,$\hfill }
\hbox to 2in {$h(\beta _{({1\over 2})})=X,$\hfill }\cr
&\hbox to 2in {$h(\alpha _{(1)})=Xe^{-2i\xi Q},$\hfill }
\hbox to 2in {$h(\beta _{(1)})=Xe^{-i\xi Q},$\hfill }\cr
&\hbox to 2in {$h(\alpha _{({3\over 2})})=Xe^{-3i\xi Q} ,$\hfill }
\hbox to 2in {$h(\beta _{({3\over 2})})=Xe^{-2i\xi Q},$\hfill }\cr
&\ldots \cr
}\eqn\ab$$
Note that
$h(\alpha _{(j)})h(\beta _{(j)})$
is independent of $j.$ This is because of the superselection rule
for the combined flux. The path $\alpha \beta $ surrounds both
vortices and can be thought of as measuring the flux at spatial
infinity.

When $(\xi/2\pi )$ is irrational, the sequence never repeats itself.
In this case, the classical configuration space is simply the
previously discussed Riemannian surface with $M=\infty .$ When
$(\xi/2\pi )$ is rational, there is a distinction between fractions
with even denominators and fractions with odd denominators. When
$(\xi/2\pi )$ has an even denominator, an even number of exchanges
(or braid operations) restores the original flux state. Thus paths
related by an odd number of exchanges will always produce a different
flux state, and should not be included in the path integral sum. The
path integral for a process involves only ``direct" paths---that is,
paths that take $A$ to $A^\prime $ and $B$ to $B^\prime $---%
and does not involve any ``exchange" paths---that is, paths that take $A$
to $B^\prime $ and $B$ to $A^\prime .$ By contrast, when $(\xi/2\pi )$
has an odd denominator, an odd number of exchanges restores the
original flux state. Thus the path integral includes both ``direct"
and ``exchange" contributions. For $(\xi/2\pi )$ irrational, the path
integral includes only one homotopy class of paths, because all homotopically
inequivalent paths result in different final flux states. However,
for certain large winding numbers the distinction becomes
arbitrarily hard to discern.

The discussion so far has been simplified because it has ignored the
zero mode and its coupling to the massless degrees of freedom
that exist because there is a continuous unbroken symmetry.
A flux state with
$h(\alpha )=Xe^{i\theta Q}$
and
$h(\beta )=Xe^{i[\theta +\xi ]Q},$
which we shall denote as
$$ \vert \theta >\otimes \vert (\theta +\xi )>\eqn\ad $$
is a superposition of total charge/flux eigenstates
$\vert Q,\xi >,$
with every charge occurring in the superposition with
the same amplitude. The bases are related by the expression
$$\vert Q,\xi >=\int _0^{2\pi }{d\theta \over \sqrt{2\pi }}
e^{i\theta Q}\vert \theta  >\otimes
\vert (\theta +\xi )>.\eqn\ae$$
Let $R$ be the vortex-vortex separation. For total charge/flux
eigenstates, there is a classical potential of the form
$-Q^2{\rm ln}[R].$ The repulsion is due to the fact that
a pair with Cheshire charge lowers its self-energy by separating,
and thus further delocalizing the Cheshire charge.

A realistic discussion of vortex-vortex scattering should take into
account both effects. In the weak coupling limit, the `Coulomb'
interaction diminishes, while the strength of the purely
quantum-mechanical holonomy effect remains constant.

\vskip 15pt
\chapter{Conclusion}

This paper has explored several of the Aharonov-Bohm phenomena which
would occur if the simplest realistic cosmic strings exist.  Among
possibilities we find would not occur is AB catalysis of quark-
lepton transitions for trajectories passing outside the core of a
string.  Nevertheless, there would be interesting and observable
color interactions which would produce hadronic excitations and
perhaps characteristic hadronic jets resulting from hadron-string
collisions. These interactions represent perhaps the first realistic
example and consequence of Alice strings, and they turn out to be non-Abelian
Alice strings rather than the Abelian ones which introduced the concept.

The general
formalism for AB scattering has been applied to a beautiful formal problem, the
scattering of straight parallel strings on each other, so that the motion is
restricted to the two space dimensions perpendicular to the string
axes, and thus is really a problem in 2+1 dimensional quantum mechanics.
Because of its entirely topological nature, the flux-flux scattering
discussed in section 4 can be interpreted as a statistical
interaction of a rather exotic sort. It is well known that in two
spatial dimensions (as contrasted with three or more spatial dimensions)
there are more possible types of statistics because the homotopy
classes for ways of interchanging $n$ particles are described by the infinite
braid group $A_n$ rather than the much smaller finite symmetric group $S^n.$
``Anyons" have exotic statistics described by one-dimensional unitary
representations of $A_n.$\refmark{\anyons }  Anyonic statistics are Abelian in
the sense that  the non-Abelian structure of $A_n$ is not probed---that is,
$g_1g_2g_1^{-1}g_2^{-1}$ is always mapped into a trivial phase.
This is no longer the case for the non-Abelian vortices discussed here.
It would be very interesting to find examples of
systems in a two-dimensional condensed matter context
with elementary excitations which obey similar non-Abelian
statistics.

An intriguing subject which still presents some open questions is the
issue of string excitations corresponding to localization of charge on
the string.  Depending on the energy cost of such excitations the transition
from quark to lepton for fermions which penetrate the string, a process which
can have large cross section growing as a fractional power of the de Broglie
wavelength, could be accomplished by deposit of the lost charges on the string
core. Estimates of the charge deposit energy would be a worthwhile subject for
future study.

Cosmic strings in themselves are an example of the Aharonov-Bohm effect, in
the sense that the continuity and covariant constancy of the Higgs fields
surrounding the string constrains the allowed fluxes carried by the string.
Thus the two subjects are intimately connected, and very likely there remain
new links between them to be explored in the future.

\vskip 15pt

\leftline{\bf Acknowledgements.} We thank
Robert Brandenberger, Anne Davis, Chung-Pei Ma,
Leandros Perivolaropolous, and Frank Wilczek
for useful discussions.  This work was supported in part by
Department of Energy grant DE-FG02-90ER40542
and in part by the National Science Foundation
under grant PHY 9309888.

\refout

\vskip 15pt
\leftline{FIGURE CAPTIONS}

{\bf Figure 1. Describing a Classical Vortex Configuration.}
Four vortices $V_1,\ V_2,\ V_3,\ V_4$ are represented
in a constant timeslice. A basis of paths $C_1,\ C_2,\ C_3,\ C_4$
starting and ending around $x_0$ is indicated. The paths are used
to define the fluxes carried by the vortices.  Each path $C_j$ winds
around the vortex $V_j$ exactly once in the counterclockwise direction
without winding around any of the other vortices.

{\bf Figure 2. Braid Dependence.}
In this figure it is shown that the winding number of a
path does not determine the flux completely. The two paths $\alpha $
and $\alpha '$ have the same winding numbers around the vortices
$A$ and $B.$ However, when the vortices $A$ and $B$ carry non-Abelian
fluxes that do not commute, the fluxes measured with respect to $\alpha $
and $\alpha '$ differ. Since $\alpha '=\beta \alpha \beta ^{-1},$
$h(\alpha ')=h(\beta )h(\alpha )h(\beta )^{-1}.$
Thus $h(\alpha ')=h(\alpha )$ if and only if
$[h(\alpha ),~ h(\beta )]= h(\alpha )h(\beta )h(\alpha )^{-1}h(\beta )^{-1}=e.$

{\bf Figure 3. Vortex-Vortex Scattering.}
In (a) is indicated a basis for two vortices
positioned at $A$ and $B.$ For simplicity, we imagine that vortex $B$ is
held fixed while vortex $A$ is allowed to propagate quantum mechanically.
We consider the propagation of vortex $A$ to its new position $A',$
as indicated in (b). In (b) are shown two homotopically inequivalent
paths, path 1 and path 2 from $A$ to $A'.$ (Actually, there is an infinite
number of homotopy classes of paths from $A$ to $A'.$)
In (d) is indicated a basis for the final state. If path 1 is taken
$\alpha _f=\alpha_1.$ If path 2 is taken $\alpha _f=\alpha _2.$

{\bf Figure 4. Effect of Moving One Vortex Around Another Vortex.}

{\bf Figure 5. Contour for Calculating Exact Solution.}

{\bf Figure 6. Saddle Points and Deformations of Contours.}

{\bf Figure 7. The Braid Operation.} The braid operation ${\cal R}$
is illustrated in (a). Fig. (b) indicates a basis. Fig. (c)
indicates what happens to basis path defined in (b) when the
${\cal R}$ operation is reversed, dragging the paths back.
$\alpha _{(0)}$ ias dragged back to $\alpha _{(1/2)}$
and $\beta _{(0)}$ is dragged back to $\beta _{(1/2)}$
as the ${\cal R}$ operation is reversed.

{\bf Figure 8. Moving a quark through a loop of colored Alice string.}
Initially, the colored Alice string is uncharged and the $Q\bar Q$
pair is in a singlet state. The quark is taken through the loop
and then back to its original position, while the antiquark remains
fixed. Because of the transfer of Cheshire charge, the final state is a
superposition of a state for which no charge transfer has taken place,
with both the loop and the $Q\bar Q$ pair in color singlet states,
and a state with the both the $Q\bar Q$ pair and the loop in color
octet state, indicating the transfer of Cheshire charge.

\end